%% file: ms.tex
\documentclass[useAMS,usenatbib]{mn2e}

\usepackage{amssymb}
\usepackage[pdftex]{graphicx}
\newcommand{\ve}[1]{\mathbf{#1}}

\newcommand{\eqref}[1]{\ref{#1}}

\newcommand{\eg}{\emph{e.g.\ }}
\newcommand{\cf}{\emph{cf.\ }}

\title[Non-linear structure topology in the 2dFGRS]{Topology of non-linear structure in the 2dF Galaxy Redshift Survey}
\author[J.~B.~James, M.~Colless, G.~F.~Lewis \& J.~A.~Peacock]{J. Berian James$^{1,3}$\thanks{Lead author contact:
jbj@roe.ac.uk}, Matthew Colless$^{2}$, Geraint F. Lewis$^{3}$
and John A. Peacock$^{1}$\\
$^{1}$Institute for Astronomy, Royal Observatory, Blackford Hill, Edinburgh EH9 3HJ, United Kingdom\\
$^{2}$Anglo-Australian Observatory, PO Box 296, Epping NSW 1710, Australia\\
$^{3}$School of Physics, The University of Sydney NSW 2006, Australia}

\begin{document}

\date{Accepted 2008 December 7. Received 2008 September 29; in original form 2008 February 24}
\volume{394}
\pagerange{454--466} \pubyear{2009}

\maketitle

\label{firstpage}

\begin{abstract}
We study the evolution of non-linear structure as a function of scale in samples from the 2dF Galaxy Redshift Survey, constituting over 221 000 galaxies at a median redshift of $z=0.11$. The two flux-limited galaxy samples, located near the southern galactic pole and the galactic equator, are smoothed with Gaussian filters of width ranging from $5$ to $8$ $h^{-1}$Mpc to produce a continuous galaxy density field. The topological genus statistic is used to measure the relative abundance of overdense clusters to void regions at each scale; these results are compared to the predictions of analytic theory, in the form of the genus statistic for i) the linear regime case of a Gaussian random field; and ii) a first-order perturbative expansion of the weakly non-linear evolved field. The measurements demonstrate a statistically significant detection of an asymmetry in the genus statistic between regions corresponding to low- and high-density volumes of the universe. We attribute the asymmetry to the non-linear effects of gravitational evolution and biased galaxy formation, and demonstrate that these effects evolve as a function of scale. We find that neither analytic prescription satisfactorily reproduces the measurements, though the weakly non-linear theory yields substantially better results in some cases, and we discuss the potential explanations for this result.
\end{abstract}


\section{Introduction}\label{sec:intro}

The large-scale structure of the cosmological density field contains important information about the initial conditions and evolution of the Universe. This structure evolves solely under the influence of gravity and its properties are a strong function of scale. Specifically, on the largest scales the morphology of the structure will reflect its primordial shape, expected to be that of nearly scale-invariant Gaussian random density perturbations~\citep{BST,1993ppc..book.....P,Peacock99}, generated by the scalar field, or fields, responsible for cosmological inflation~\citep{Guth81,PhysRevLett.48.1220,Liddle00,2005NewAR..49...35L}. That this description accords with observation is strongly supported by galaxy redshift surveys~\citep{1989Sci...246..897G,1996ApJ...470..172S,Colless01,2000AJ....120.1579Y} and microwave background measurements~\citep{1994ApJ...436..423B,2006astro.ph..3449S}, as well as the methods used to obtain information about morphology of the large-scale structure~\citep{Gott89,Melott89,1992ApJ...385...26G}.

Indeed, the tools for measuring these properties---the topology of the large-scale structure---are beginning to find wider application. It is more than 20 years since the publication of the seminal work of \citet{GMD}, and the field has matured substantially. The initial question---whether the topology of the cosmological density field is consistent with inflation---has been answered in the affirmative through repeated analysis~\citep{Gott89,1994ApJ...420..525V,Canavezes98,Hikage02,HoyleSDSS,Hikage03,2005ApJ...633...11P,2007MNRAS.375..128J} and studies of the departure from this as a result of non-linear gravitational evolution, rather than as an initial condition, are an exciting new direction for observational studies (\cf \citealt{Melott88} for pioneering work with simulations).

The classic topological statistic is the genus number of constant-density surfaces through the field, with the surfaces ordered so as to excise increasingly larger volumes. Each genus number is a measure of the connectedness of the structure at a particular density; the resulting locus, called the genus curve, is genus number as a function of density relative to the median. The most important qualitative feature of this curve, for a Gaussian random field, is symmetry about this median density---the high- and low-density regions of the immediate post-inflation Universe are equivalent and interlocking, giving rise to a so-called \emph{sponge topology}. With regard to the configuration of regions above and below median density, this is one of only three possibilities: it is the middle ground between the universe evoked by~\citet{1974ApJ...187..425P}, a connected void populated with high-density clusters---a \emph{meatball topology}---on the one hand, and the single web of high-density structure punctured by voids---a \emph{swiss-cheese topology}---on the other. All three are idealisations: a spectrum runs smoothly from one extreme to the other, leading to the more common characterisation in terms of meatball- and swiss-cheese-\emph{shift} away from the norm.
 
The genus curve has proven to be an especially valuable measure, as it can be predicted analytically for the case of the Gaussian random density field~\citep{1970Ap......6..320D,1981grf..book.....A,BBKS,Hamilton86}. That the topology of structure will be scale-dependent should follow from the predictions of inflationary models together with observations of the local Universe, where isolated high-density structures predominate. This will produce a transition in the genus curve from symmetry at (sufficiently) large scales to a strong asymmetry in the non-linear regime of structure growth on small-scale and at late times. Na\"ively, such departures would be expected to arise at about the correlation length, where $\delta\rho/\rho\sim1$ and modes of different sizes are beginning to interact. However, the form of the genus curve for a general field that does not obey Gaussian statistics has resisted analytical description, despite significant theoretical headway for specific cases of non-Gaussian fields~\citep{1988PASP..100.1343H,1996ApJ...463..409M,2007arXiv0711.3603H} and second-order perturbation expansions of the Gaussian result~\citep[based on the methods of][]{1994ApJ...420...44K,1995ApJ...442...39J} into the weakly non-linear regime~\citep{1994ApJ...434L..43M,Matsubara96,1997ApJS..110..177S}. The latter case is especially salient and \citet{2003ApJ...584....1M} provides a key conclusion: that the effect of the weakly non-linear evolution on the genus curve is relatively modest.

It is against these expectations that the data must be tested. The weakly non-linear approximation will be observed to break down on sufficiently small scales and the behaviour of the genus curves throughout and beyond the regime in which it is valid provides higher-order information on the evolution of the density field than would be garnered with correlation functions or power spectra. Such a measurement requires a survey of a large volume with statistical properties that are well understood. The considerable sophistication of the statistical methods that have been brought to bear on galaxy redshift surveys since the turn of the century provide an opportunity to address this topic directly.

In this paper, we use the final data release of the Two-degree Field Galaxy Redshift Survey (2dFGRS) to make measurements of the genus statistic on the cosmological density field on scales ranging from 5 to 8 $h^{-1}$Mpc. This builds upon the analysis of statistical properties of the galaxy distribution with this data set, including the power spectrum~\citep{Cole05}, void probability function~\citep{Croton04a}, the two-point and higher-order correlation functions~\citep{Hawkins03,Croton04b} and the genus statistic measured in the linear regime as a test of the inflationary hypothesis~\citep{2007MNRAS.375..128J}. Section~\ref{sec:survey} describes how the density field is reconstructed from the galaxy survey and the restrictions this places on subsequent measurements. The genus measurement itself is the subject of Section~\ref{sec:genus}, as are the physical meanings of the different qualitative results that may be obtained. Section~\ref{sec:results} presents the results and method of error analysis for genus measurements of the 2dFGRS data, in which we test for the presence of a shift in the genus curve and discuss the implications of these results. Throughout, we use a cosmology in which $\Omega_m = 0.27$, $\Omega_\Lambda = 0.73$ and $H_0=100h$ km sec$^{-1}$ Mpc$^{-1}$, with $h=0.72$.

\section[]{Samples from the 2dFGRS}\label{sec:survey}
At completion, the Two-degree Field Galaxy Redshift Survey~\citep{Colless01} measured spectroscopic redshifts for 221 000 galaxies in regions of the sky well separated to ensure statistical independence. With a limiting magnitude of $19.45$ in the $b_\mathrm{J}$ band~\citep{1982PASP...94..742B}, a median redshift of $\bar{z} = 0.11$ is obtained with close-to-uniform sampling across the sky. The survey comprises two regions located near the South Galactic Pole (SGP) and slightly to the north of the Galactic equator (NGP), along with a number of smaller random fields that will not be used for the analysis in this paper, giving a total survey area of over $2000$ deg$^2$. The larger SGP region is a strip of approximately $90^\circ\times15^\circ$ ($21^\mathrm{h} 40^\mathrm{m} < \alpha < 03^\mathrm{h} 40^\mathrm{m}$, $-37.\!\!^\circ5 < \delta < -22.\!\!^\circ5$) in which the redshifts of 115 492 galaxies are avilable for use in a statistically complete sample; the NGP strip is approximately $75^\circ\times10^\circ$ ($09^\mathrm{h} 50^\mathrm{m} < \alpha < 14^\mathrm{h} 50^\mathrm{m}$, $-7.\!\!^\circ5 < \delta < +2.\!\!^\circ5$) and contains 79 696 galaxies.

\subsection{Selection effects}
The data reduction includes extinction and $k+e$ corrections, as well as masking of the survey area to account for several distinct selection effects: i) variation in magnitude limit between the two-degree fields that tile the full survey regions; ii) variation between fields in the fraction of objects with high-quality spectroscopic redshifts; and iii) variations due to magnitude-dependent redshift completeness. These corrections are performed with the publicly available software of Peder Norberg and Shaun Cole, documented by~\citet{Norberg02}. Information on the software, and the code itself, is provided on the 2dFGRS website at \texttt{http://www2.aao.gov.au/2dFGRS/}.

Correcting for these effects addresses statistical incompleteness across right ascension and declination. When generating samples that maintain a constant number density with redshift, the effect of decrease in absolute magnitude limit with distance must be addressed. When the luminosity function of the population of observed galaxies is well understood, it is possible to estimate the fraction of galaxies---as a function of magnitude and redshift---that go unobserved as a result. The inverse of this number can be used to weight the galaxies that \emph{are} seen, which serve as markers for the hidden population.

The combination of all statistical completeness effects is estimated with a single selection function, giving a weight to each galaxy. Because such a weighting scheme is useful to many different analyses of a single data set, the 2dFGRS selection function was computed by \citet{Norberg02}; they describe the estimators used to produce the weightings as well as other salient aspects of the $b_\mathrm{J}$-band luminosity function. Galaxy samples produced in this manner are \emph{flux-limited} and retain all objects in the original catalogue at the expense of uncertainty over the exact values of the weighting and limitations imposed by the estimation.

An alternative to using a weighted galaxy sample is to remove the redshift-dependent effect by drawing a sample comprising just those objects that would be visible at the edge of the sample region, given the magnitude-limit of the survey~\citep[\emph{e.g.}][]{MartinezSaar}; such \emph{volume-limited} samples are the more common way of making a measurement of the topological genus from a redshift survey\citep[e.g.,][]{Gott89,Canavezes98,Hoyle2dF,2005ApJ...633...11P,2006astro.ph.10762G}; \citet{1994ApJ...420..525V} is a notable counterexample. Flux-limited samples of the 2dFGRS were used by \citet{2007MNRAS.375..128J}, who demonstrated that i) the choice of flux-limited catalogues reduced the noise in the measurement without introducing spurious effects; and ii) for the purposes of topological analysis, the selection function of the 2dFGRS is best restricted to the region $z < 0.2$. 

\subsection{Sample constraints}\label{sec:constraints}
The choice of the maximum redshift to which a galaxy sample should extend depends on the properties of the underlying catalogue and the scale that is to be studied. This length scale is imposed through the smoothing process, where it is identified with the characteristic width of the filter. The most restrictive requirement is that the smoothing length, $\lambda$, exceed the average nearest-neighbour separation of galaxies in the sample volume. The latter quantity derives from the mean number density of the survey, $\bar{n}$; it is proportional to the cube root of the volume per object, $\bar{n}^{-1/3}$. For a Poisson point process, the mean nearest neighbour separation is
\begin{equation}
\bar{\ell}_\textrm{nn} = \frac{\bar{n}^{1/3}}{\sqrt{\pi}} = \left(\frac{V}{\pi^{3/2}N}\right)^{1/3},
\end{equation}
where $V$ is the volume of the survey and $N$ is the number of galaxies. However, the cosmological density field is not a continuous Poisson field, nor is the drawing of points from the background an unbiased process, so galaxies in redshift surveys should be assumed \emph{not} to be a Poisson process, though the qualitative form of the relationship between $\bar{\ell}_\mathrm{nn}$ and $\bar{n}$ should not change. A safe choice of smoothing length for a Gaussian filter is
\begin{equation}
\lambda > \bar{\ell}_\mathrm{nn} \approx \left(\frac{V}{N}\right)^{1/3};
\end{equation}
this prescription has been used previously with success by \citet{1994ApJ...420..525V}. 
The average number density is a function of the size of the survey; for a survey through a fixed window in right ascension and declination, $\bar{n}$ is a function of $z_\mathrm{max}$ only. In this way, the scale to be probed influences the choice of maximum redshift used when drawing samples from the 2dFGRS catalogue.

Another relationship between $\lambda$ and $z_\mathrm{max}$ arises from considering the number of resolution elements in the smoothed sample. A high number of resolution elements is desirable in order to recover a statistically significant signal, though there is no critical threshold. The number of resolution elements is
\begin{equation}
N_\mathrm{res} = \frac{V_\mathrm{survey}}{V_\mathrm{kernel}} = \frac{V}{\pi^{3/2}\lambda^3} \Rightarrow \lambda = \left(\frac{V}{\pi^{3/2}N_\mathrm{res}}\right)^{1/3}.
\end{equation}
A value of $N_\mathrm{res} = 10^3$ has been selected as indicative of a measurement of reasonable significance. 

\begin{figure*}
\centering
\includegraphics[width=185mm]{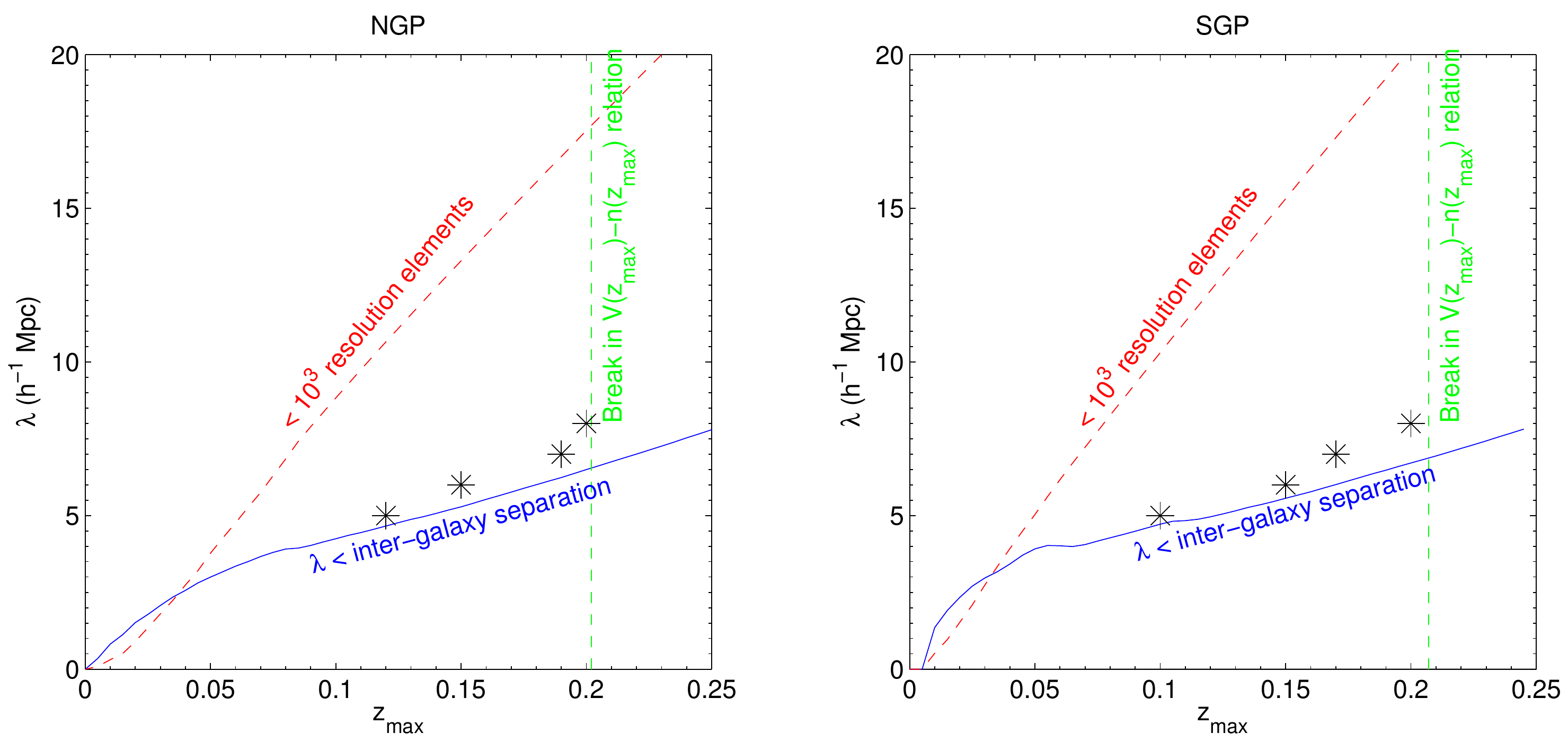}
\caption{Constraint diagrams for the point samples drawn from the two 2dFGRS catalogues. Considerations of the selection function (dashed green), spatial number density (solid blue) and resolution (dashed red) restrict values of $z_\mathrm{max}$ and $\lambda$ to the central triangular region. The constraint on the number of resolution elements is only indicative ($N_\mathrm{res}=10^3$); it is desirable to draw samples as close the solid line as possible, so as to maximise the signal-to-noise of the measurement while avoiding spurious signal from sparse points being treated as isolated clusters. The points indicate the choice of $\lambda$ and $z_\mathrm{max}$ employed in the samples used in this work.}\label{fig:constraint}
\end{figure*}

As has been stated, the choice of $z_\mathrm{max}$ is also restricted on account of the survey selection function; Figure 2 of \citet{James05} shows a break in the slope of the cumulative number density as a function of redshift at $z_\mathrm{max}=0.2$, leading to a natural choice for cutting off the maximum redshift at which an object should be included in this measurement. This defines a relation within the $z_\mathrm{max}$--$\lambda$ plane, albeit one that is independent of $\lambda$. This plane, along with the the constraints these three relations provide, is shown in Fig.~\ref{fig:constraint}.

Point samples are drawn from the NGP catalogue out to those values of $z_\mathrm{max}$ that will allow smoothing on scales of $\lambda = 5$ to $8 h^{-1}$Mpc in $1 h^{-1}$Mpc increments. The NGP catalogue has higher number density at low redshift, as it probes a smaller volume; the measurements in samples from the SGP catalogue will be correspondingly stronger and are also made above $5 h^{-1}$Mpc. Table~\ref{table:samples} shows the salient parameters of each of the point samples drawn from the NGP and SGP slices.
\begin{table}
\centering
\begin{tabular}{|c|c|c|c|c|c|c|}
\hline \hline
NGP & & & & & \\\hline
$\lambda$ & $z_{max}$ & $V$ ($/ 10^6$)& $N$ & $1/\bar{n}$ & res & $p_{(\%)}$ \\\hline
$5$ & 0.12 & 6.652 & 48892 & 136.0 & 0.20 & 11.44 \\ 
$6$ & 0.15 & 12.98 & 63431 & 204.6 & 0.25 & 10.31 \\
$7$ & 0.19 & 25.74 & 73897 & 348.3 & 0.29 &  9.33 \\
$8$ & 0.20 & 29.86 & 75268 & 396.7 & 0.25 & 10.18 \\
\hline 
\hline
SGP & & & & & & \\\hline
$5$ & 0.10 & 5.733 & 42441 & 135.1 & 0.20 & 10.94 \\
$6$ & 0.15 & 19.19 & 82399 & 232.9 & 0.25 &  7.33 \\
$7$ & 0.17 & 27.66 & 92893 & 297.7 & 0.29 &  7.45 \\
$8$ & 0.20 & 44.11 & 103616 & 425.7 & 0.25 &  7.68 \\
\hline \hline
\end{tabular}\caption{Description of the point samples used in the genus measurement: for a given smoothing scale ($\lambda$, in $h^{-1}$Mpc), a value for the redshift limit ($z_\mathrm{max}$) of the sample is chosen given the constraints of Figure~\ref{fig:constraint}; this defines the volume ($V$, in cubic comoving megaparsecs), number of galaxies ($N$) and number density ($\bar{n}$) of the sample and the resolution of the array cells (in units of $\lambda$). The sum of galaxy weights (the `power', a proxy for mass) is not conserved by the smoothing process, which pushes power outside the edges of the survey region. By normalising the smoothed galaxy distribution to a smoothed constant field, no systematic effects are introduced by this process, though the loss of power ($p_{\%}$) reduces the quality of the signal.}\label{table:samples}
\end{table}

\subsection{Sample construction}\label{sec:construction}
The flux-limited galaxy samples from the NGP and SGP catalogues include all objects out to a maximum redshift given the constraints above. Initial input from the 2dFGRS catalogue takes the form of a list of redshift, right-ascension and declination values, as well as a weight for each galaxy corresponding to the inverse of the completeness. The redshift is converted to comoving distance~\citep[\emph{e.g.~}][]{Hogg99} by numerically integrating
\begin{equation}
D_c = \frac{c}{H_0}\int_0^z\frac{dz^\prime}{\sqrt{\Omega_m\left(1+z\right)^3 + \left(1-\Omega_m-\Omega_\Lambda\right)\left(1+z\right)^2 + \Omega_\Lambda}};
\end{equation}
the distance and angular co\"ordinates are converted to Cartesian co\"ordinates, which are then rotated through the $y$-axis (corresponding to declination) so that the volume of the bounding box is minimised. 

\begin{figure*}
  \centering
  \includegraphics[width=0.32\linewidth]{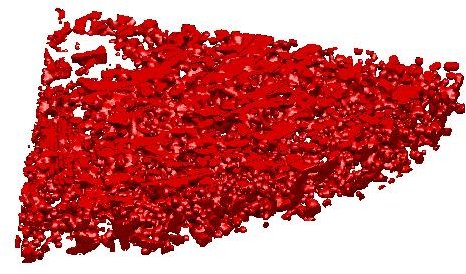}
  \includegraphics[width=0.32\linewidth]{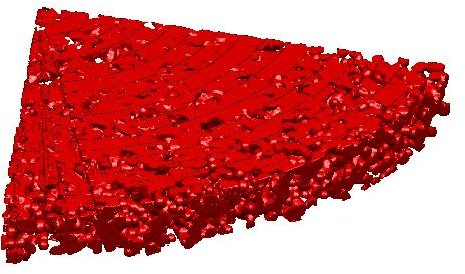}
  \includegraphics[width=0.32\linewidth]{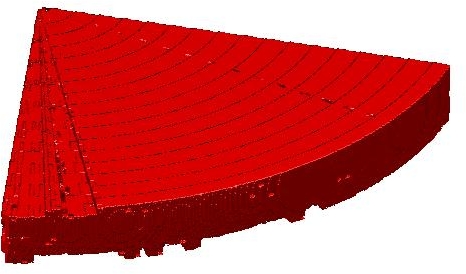}
  \includegraphics[width=\linewidth]{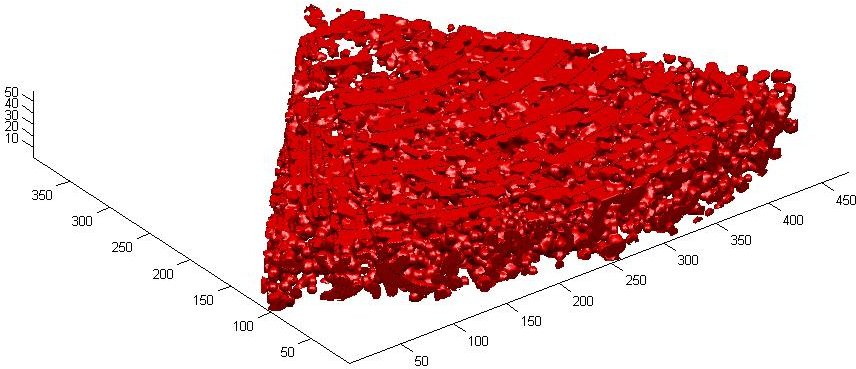}
  \caption{Surfaces of constant density through the cosmological density field of the NGP strip, extended out to a redshift of 0.2. The field is recreated by smoothing the statistically-complete point catalogue with a Gaussian filter, in this case of width 4 Mpc, and subsequently embedding it within the data array, as demonstrated in the lower panel. The images in the upper panel show the changing topology of the surfaces as different volume are excised: shown are surfaces containing (from top left) 33\%, 66\% and 100\% of the volume within the sample. The single lower panel excises 50\% of the field---corresponding to the median density---showing the geometry of the NGP region within the $394\times475\times60$ array along with the effect the survey mask. At the high-redshift end of the wedge, many isolated spherical surfaces are apparent, indicating that $z_\mathrm{max}=0.2$ is too large given the smoothing length; \emph{cf.} Figure~\ref{fig:constraint}.} \label{fig:ngp_wedges}
\end{figure*}

\begin{figure*}
  \centering
  \includegraphics[width=0.32\linewidth]{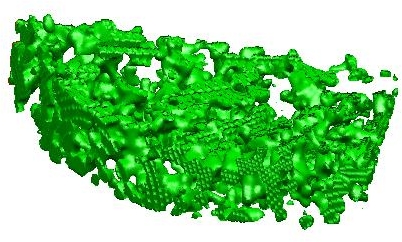}
  \includegraphics[width=0.32\linewidth]{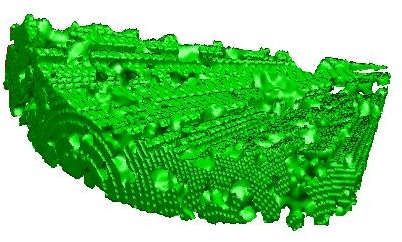}
  \includegraphics[width=0.32\linewidth]{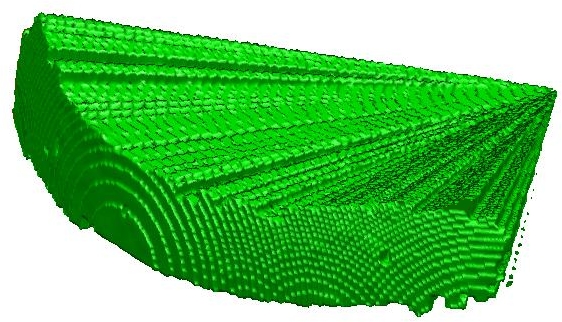}
  \includegraphics[width=\linewidth]{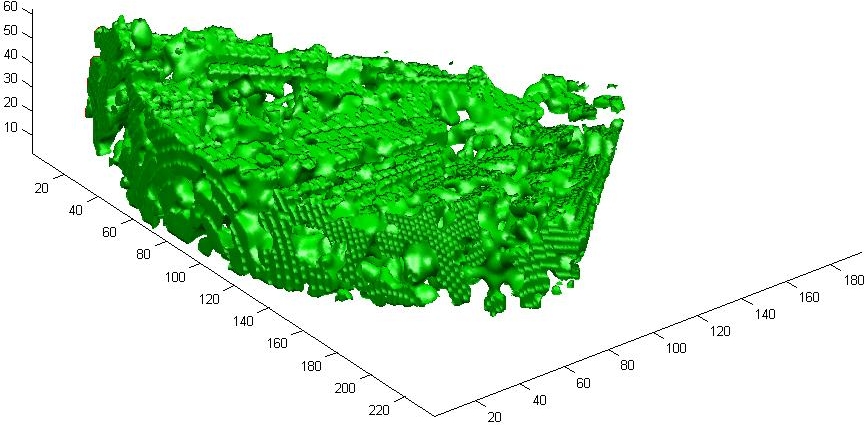}
  \caption{Isosurfaces of the SGP strip, extending to a redshift of 0.2, smoothing with a Gaussian of width $\lambda=8h^{-1}$ Mpc. The upper panels again show the surfaces containing (from top left) 33\%, 66\% and 100\% of the volume within the sample. Of note is the significantly smoother structures generated by the larger smoothing kernel; the topology is very close to that of a Gaussian random field.} \label{fig:sgp_wedges}
\end{figure*}

For the purpose of the smoothing and the genus measurement, the galaxies are represented within a three-dimensional array, binned using a nearest-grid-point algorithm, with resolution set to be well below the smoothing length; using bins of size $2 h^{-1}$Mpc is deemed adequate for smoothing of $5 h^{-1}$Mpc or more. Figures \ref{fig:ngp_wedges} and \ref{fig:sgp_wedges} show the NGP and SGP slices embedded in this data structure. Cells outside the survey region must be flagged as such, rather than empty; they are given negative values. 

However, by discretising the spatial set of galaxy locations, those regions inside the slice removed during masking must be also be flagged. If this is not done, these cells will act as false voids, leading to a Swiss-cheese signature in the genus measurement. Masking for the survey regions based on the composite survey completeness maps is applied by flagging regions of the $\alpha$-$\delta$ plane within and around the NGP and SGP slices that are not part of the survey.
The flagged regions are projected along the line-of-sight to define a pencil beam of flagged cells within the data array. The incursion of these cut-outs around the edge of the slices (and the SGP slice in particular) is apparent in Figures~\ref{fig:ngp_wedges} and~\ref{fig:sgp_wedges}.

The correctly masked data array is a three-dimensional structure with a flagged (negative) value in cells outside the survey bounds, as well as in masked regions within the bounds. The remaining cells have positive value and represent the weighted galaxy sample.

\subsection{Sample Smoothing}\label{sec:smoothing}
To retrieve an estimate of the cosmological density field, the galaxy sample is smoothed through a Gaussian window, though spherical top-hat~\citep{Hamilton86} and wavelet techniques~\citep{Martinez05} have also been considered for the purposes of the genus measurement. The width of the smoothing kernel acts to select a scale, in Mpc, of the features of the field to be emphasised. As a result, by using kernels of a range of widths, the measurement is made across a range of scales. The appropriate range is governed by the statistical constraints of the sample, developed in Section~\ref{sec:constraints}, and by the aim of studying the non-linear regime. The transition to scales where the evolution of structure becomes non-linear, where modes of different scales become correlated, occurs at a characteristic length where the RMS amplitude of density perturbations is of the same order as the background, i.e.\ where $\delta\rho/\rho\sim1$. Measurements of the variance of this fractional RMS amplitude between spheres of radius $8h^{-1}$ Mpc, $\sigma_8$, have converged about 0.9~\citep{Tytler04}, though disagreement persists~\citep[\emph{e.g.}][]{2006astro.ph..3449S}. Here, the same value as other clustering analyses with the 2dFGRS data set is used: $\sigma_8 = 0.89$. To study the weakly non-linear regime, therefore, is it necessary to probe around $8 h^{-1}$Mpc; based on the sampling constraints outlined in Section~\ref{sec:constraints}, kernel widths of $5$, $6$, $7$ and $8$ $h^{-1}$Mpc have been chosen for both regions. These are related to the spherical top-hat filter radius, which corresponds more precisely to the notion of physical distance, by $\lambda_\mathrm{G} = \lambda_\mathrm{T}/\sqrt{5}$; this result follows from the conservation of the filter volumes and the distinction should be kept in mind when interpreting the results.

The smoothing is carried out \emph{via} Fourier-space convolution\footnote{We use the FFTW libraries~\citep[][v3.1.2 from \texttt{http://www.fftw.org}]{FFTW05} for the discrete Fourier transform.} of the sample with a Gaussian kernel,
\begin{equation}
G(\ve{x}) = \exp\left(-\frac{\ve{x}^\prime\cdot\ve{x}^\prime}{2\lambda^2}\right)\label{eq:gaussian};
\end{equation}
as the survey regions do not completely, or even nearly, fill the data volume---the slice geometry occupies approximately one-third of a tightly-fitting rectangular prism---there will be power lost through the outer surfaces of the slice from galaxies smoothed partially into the flagged non-survey region. This signal is not recoverable and creates a relative drop in power near the slice boundaries that will, if left untreated, produce a volumetric over-abundance of low-density region, leading to a swiss-cheese signature. To avoid this, the density must be artificially raised in a manner that does not disturb the topological measurement. 

\citet{1993ApJS...86....1M} have demonstrated a method that correct for this systematic near the boundaries: create a three-dimensional array mirroring the geometry of the sample, but filled with a constant density field (say of value $\rho=1$) rather than sparsely-separated galaxies. Smoothing this constant data array mimics the power loss through the surfaces of the survey boundary; dividing the smoothed 2dFGRS data array cell-wise by the smoothed constant data array will remove this effect. Furthermore, they demonstrate that the only systematic effect induced by this process is a slight increase in the amplitude of the measured genus curve, a correction proportional to the ratio of surface area to volume that is negligible in this context; moreover, there is no change to the symmetry properties of the curve that are studied in this work. A more important penalty is the decrease in signal---when $\lambda$ is even a moderate percentage of the survey size, the fraction of signal lost can be quite large; these values are the final column of Table~\ref{table:samples}.

\section{Topological Analysis}\label{sec:genus}

The morphology of the galaxy distribution has long been known to convey information from all orders of the correlation function, albeit in a way that is yet to be completely elucidated. Unlike geometric measures of structure, morphological statistics concentrate on properties such as `connectedness' and global curvature rather than orientation or size. The most famous of these, the Minkowski functionals~\citep{Mecke94,Kerscher96,Schmalzing97}, give a concise description of the density field up to topological equivalence; for two-dimensional surfaces embedded in three-dimensional space, they correspond to the enclosed volume, area, integrated mean curvature and integrated Gaussian curvature of the surface.

The latter is proportional to the number of holes through the surface, called the genus number; informally
\begin{equation} g = \textrm{number of holes} - \textrm{number of isolated regions} + 1. \end{equation}
When a surface is constructed through a three-dimensional field on the basis of a physical quantity, the genus number measures how connected (when $g>0$) or disjoint ($g<0$) regions demarcated by this quantity are. The quantity of interest to cosmologists is mass density; for instance, if the surface is drawn at the critical density for spherical collapse, the genus number will measure how inter-linked such regions are relative to the background.

The numerical calculation of the genus number is done indirectly by computing the integrated curvature for surfaces through an array containing the smoothed density field. The algorithm of~\citet{WGM} marks regions as either above or below a critical density and defines the surface as the boundary between them. The relationship between the genus number and integrated Gaussian curvature (on a smooth manifold) is the Gauss-Bonnet theorem
\begin{equation} \int_S K dA + \int_{\partial S} k ds = 4\pi(1-g), \end{equation}
where $K$ is the Gaussian curvature and $k$ is the curvature on the boundary of the surface; the second integral vanishes whenever the surface is closed. Because the computation of the genus number takes place in a polygonal data structure, the surface is composed of the flat sides of array cells along with the vertices and edges that join them; it is the result of an earlier theorem of Descartes~\citep{Descartes,Sasaki03} on the angle defect\footnote{\emph{Cf.\ }the more modern sounding `angle deficit'; while the description of this measurement as an application of the Gauss-Bonnet theorem---which applies specifically to smooth manifolds---is hardly inappropriate, the more general index theorem of~\citet{AtiyahSinger} is also worthy of mention.} of polyhedra that is relevant, \emph{viz.}
\begin{equation} 4\pi(1-g) = \sum\left(2\pi - \sum_i A_i\right),\end{equation} 
where $A_i$ is the angle made by each face of the cell that meets a vertex and the outer sum is evaluated over every vertex in the array. This is the quantity that must be evaluated for each cell vertex of the array containing the smoothed density field; the angle deficit matrices tabulated by~\citet{WGM} are now used to compute the genus number of the surface.

\subsection{Genus curve of embedded surfaces}

The measurement of genus number is made not just on one surface, but on a sequence of surfaces spanning the range of densities contained within the field. The characteristic density that defines each surface is chosen to excise a fraction of the volume of the field, parameterised by the parameter $\nu$,
\begin{equation} v_f(\nu) = \frac{1}{\sqrt{2\pi}}\int_\nu^\infty e^{-t^2/2}dt = \frac{1}{2}\textrm{Erf}_c\left(\frac{\nu}{\sqrt{2}}\right), \label{eq:nu}\end{equation}
where Erf$_c$ is the conjugate error function. In this way, $\nu=0$ corresponds to splitting the field into the higher- and lower-mass regions that each occupy 50\% of the volume; roughly the highest-density 15\% of the volume lies above the surface at $\nu=\pm1$, 2\% for $\nu=2$ and so on as for the usual Gaussian integral.

Defining the value of the density that splits the field into the proper volume fraction is done by reshaping the three-dimensional density field into an ordered one-dimensional array so that, as each element of the array represents a standard unit of volume, moving $p$\% of the way along the array will give the critical density corresponding to enclosing $(100-p)$\% of the volume. That is, $\rho_c$ is the $n$th element of the sorted arrays, for
\begin{equation} n = \textrm{floor}\left[(1-v_f)N\right], \end{equation}
where $N$ is the length of the ordered array.

\begin{figure}
\centering
\includegraphics[width=\linewidth]{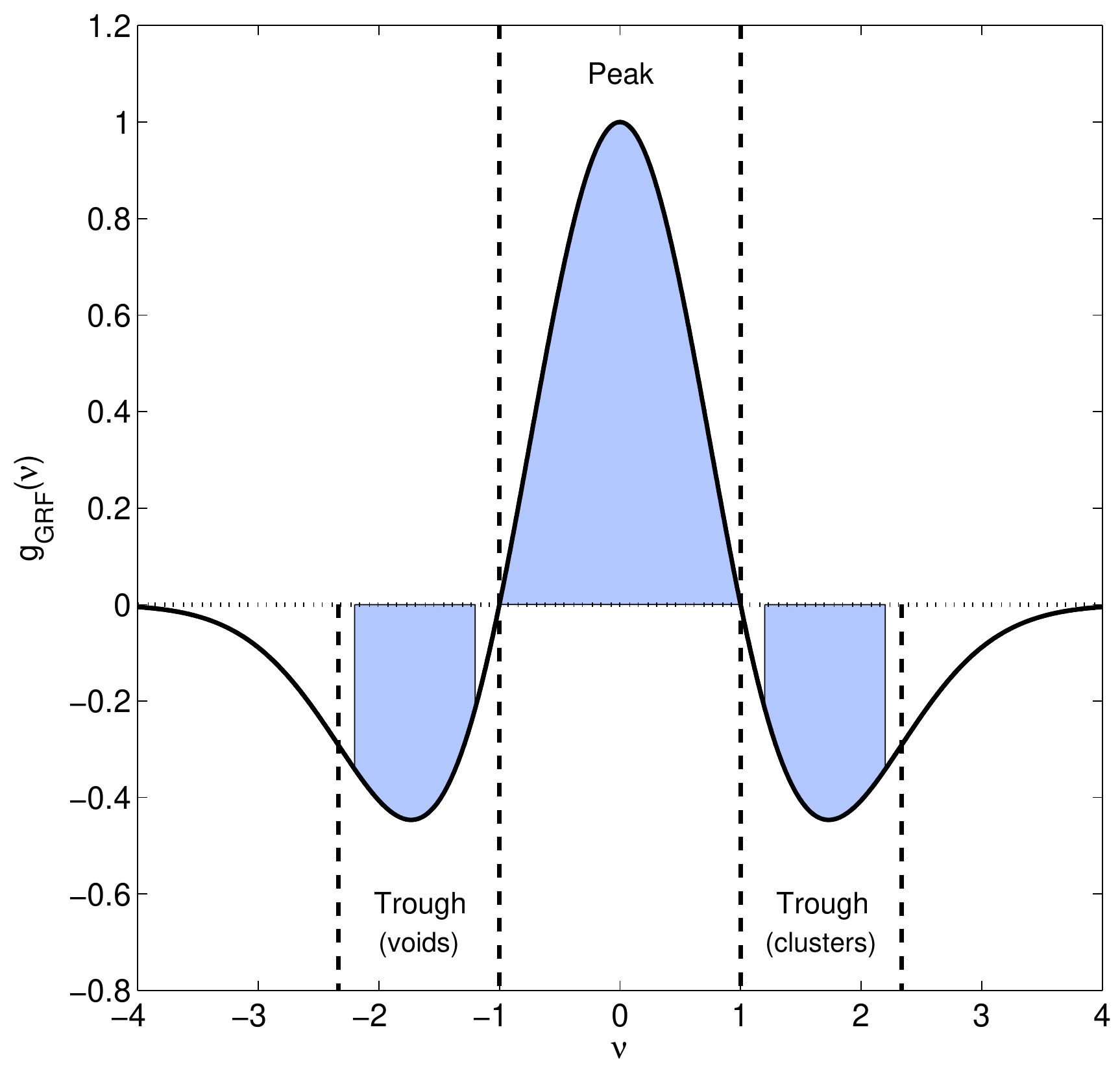}
\caption{The archetypal genus of the Gaussian random field (solid line) normalised to the peak value at $\nu=0$ and (demarcated by dashed lines) the regions of distinct topological character corresponding to the presence of isolated voids and clusters, for the troughs, and for the connected surface at median density. Interpretations of genus curve measurements focus on differences between the observed curve and that of an idealised model, such as the Gaussian case. Departures from the model are quantified using the ratio between model and measurement of the area under three regions of the curve (shaded); this is discussed in Section~\ref{sec:metastatistics}.}\label{fig:genus_curve}
\end{figure}

Excising the high- and low-density regions with this value and measuring the genus number on the intervening surface gives a curve as a function of the volume parameter $\nu$, \emph{i.e.\ }$\nu\rightarrow v_f\rightarrow \rho_c \rightarrow g$. The canonical shape of this curve for a Gaussian random field (formulated below in Equation~\ref{eq:genusgrf}) is shown in Figure~\ref{fig:genus_curve}: when $\nu$ is negative, only small disjoint regions of the field are excised, so the genus number at this value is negative; as $\nu$ increases, more disjoint regions appear and the genus number becomes even lower; eventually these regions start to join and the genus number of the surfaces increase, crossing at $\nu=-1$; the curve peaks at $\nu=0$ (median density), where the field is maximally connected; the curve then recedes symmetrically---for the case of a Gaussian field, high- and low-density regions are connected in the same way.

So, there are three domains throughout the curve: i) $|\nu|<1$, the `central peak', where the genus curve is positive, corresponding to those fractions of the volume that are more connected than isolated; ii) $1\le|\nu|\lessapprox2.5$, the `troughs', where the genus curve is negative; and iii) $|\nu|\gtrapprox2.5$, the `wings', where the value of the curve, though still negative, is trending asymptotically to zero. While the crossing point from positive to negative values of genus number serves to separate the central peak from the troughs, the point of demarcation between the troughs and the wings is less clear and somewhat more arbitrary; the points of inflection at $\nu=\pm\sqrt{3+\sqrt{6}}\approx\pm2.33$ are the obvious candidates for a rigourous division, but this is perhaps unnecessary. While such a division provides little direct insight, it has provided the implicit basis for statistical analysis of the genus curve since at least~\citet{1992ApJ...392L..51P}; the wings, as a region of especially low signal-to-noise, are often excluded from measurement, while the interplay between the central peak and the troughs guides physical interpretation.

\subsection{Modelling of the genus curve}\label{sec:models}

To properly determine the physical significance of genus measurement made on data from redshift surveys, models of the genus curve itself are required. So far, theoretical success has been restricted to the special case of the Gaussian random field---believed to accurately represent the statistical properties of the density field immediately following the end of cosmological inflation---and small departures from this case, such as would be expected under smooth gravitational evolution.

Consequently, these two descriptions are appropriate in different regimes: on large scales, at least $\lambda=10h^{-1}$ Mpc, the genus curve equation for a Gaussian field tests whether the observed density field is faithfully captured by models of the early Universe. The perturbative model applies where small departures from the Gaussian case are expected, \emph{i.e.\ } on scales below, but not much below, the divide between the linear and non-linear regimes. It is these scales that are probed by the current work.

\subsubsection{Linear regime: Gaussian random field}

Much of the early success of the genus curve as a diagnostic tool for diverging models of structure formation can be attributed to the construction of an analytic function for the special case of a Gaussian random field. As calculated by a number of independent researchers~\citep{1970Ap......6..320D,1981grf..book.....A,BBKS,Hamilton86}, the formula is
\begin{equation}
g_{\textrm{GRF}}(\nu) = A\exp\left(-\nu^2/2\right)\left(1-\nu^2\right);\label{eq:genusgrf}
\end{equation}
where the amplitude of the curve is determined by the power spectrum of the field~\citep{Hamilton86}:
\begin{equation}
A = \frac{1}{4\pi^2}\left(\frac{\langle k^2\rangle}{3}\right)^{3/2},
\end{equation} 
where
\begin{equation}
\langle k^2\rangle = \frac{\int k^2P(k)W^2(k\lambda)d^3k}{\int P(k)W^2(k\lambda)d^3k}.
\end{equation}
When the window function $W$ is a Gaussian of width $\lambda$---such as equation (\eqref{eq:gaussian})---and the power spectrum $P(k)$ a power law with spectral index $n$, the normalisation factor is
\begin{equation}
A = \frac{1}{4\pi^2\lambda^3}\left(\frac{3+n}{3}\right)^{3/2}.
\end{equation}
The significance of this result is that the curve is necessarily symmetric for general choices of power spectrum; the result is applicable even for anisotropic fields; \citet{Matsubara96} have shown that the impact of redshift-space distortions in the linear and weakly non-linear regimes is small. The high- and low-density regions of the field are treated equally with respect to the volumes they excise and to the connectedness of such regions. While the term `Gaussian random field' refers to the case where the Fourier amplitudes of the field are Rayleigh distributed and the phase distribution is uniform in the interval $[0,2\pi)$, it is the latter condition that is most salient when probing large volumes of the Universe; even fields with power distributed in a deterministic fashion (\eg $P(k)\propto k^n$) are often referred to as `Gaussian' when they satisfy the random phase condition; this amounts to an implicit invocation of the central limit theorem.

\subsubsection{Weakly non-linear regime: perturbation theory}

Correlations between phases such as those induced by non-linear gravitational evolution lead to the disruption of symmetry about the median density contour of the genus curve. The nature of this evolution has been described in a series of papers~\citep{1994ApJ...434L..43M,Matsubara96} using cosmological perturbation theory: in a Gaussian field $\delta$, all connected moments higher than $n=2$ vanish; in a non-Gaussian field satisfying a technical requirement in which higher moments are of order a specific power of the rms fluctuations of the field $\sigma$, the genus curve is given by
\begin{displaymath}
g_\textrm{WNL}(\nu) \approx -A\exp\left(-\nu^2/2\right) \times \nonumber
\end{displaymath}
\begin{equation}
\quad\left[ H_2(\nu) + \sigma\left(P_3H_3(\nu)+P_1H_1(\nu)\right)  \right],\label{eq:genuswnl}
\end{equation}
where $H_n(\nu)\equiv(-1)^n\exp\left(\nu^2/2\right)(d/d\nu)^{n}\exp\left(-\nu^2/2\right)$ are the Hermite polynomials, their coefficients $P_n$ are determined from the skewness parameters $S^{(a)}$ of the smoothed field~\citep[defined as \emph{per} equation (2.8) of][]{Matsubara96}; the approximation to the genus formula in equation (\eqref{eq:genuswnl}) arises only from discarding terms of order $\sigma^2$ and above; and the normalisation $A$ is unchanged from the linear regime. It is this equation that can be used to predict the genus curve on scales below the linear regime and it has been used in the analysis of the results. 

The genus formula for the Gaussian random field is expressible in this formalism as well, \emph{viz.}
\begin{equation}
g_\textrm{GRF} = -A\exp\left(-\nu^2/2\right)H_2(\nu),
\end{equation}
where now, with only an even Hermite polynomial present, the genus curve is symmetric about $\nu=0$---notably, this will not be the case in the weakly non-linear regime. The value of the skewness parameters, as well as the rms fluctuation in amplitude of the field, change as a function of scale. The Hermite coefficients are given simply by
\begin{equation}
P_3 =  S^{(1)}-S^{(0)},\quad P_1 = S^{(2)}-S^{(0)};
\end{equation}
over the range of scales probed in this work, the values of the skewness parameters---and hence those of the Hermite coefficients---can be fit by a cubic polynomial. Table~\ref{table:coeffs} shows the polynomial coefficients of the parameters,
\begin{table}
\centering
\begin{tabular}{|c|c|c|c|c|}
\hline \hline
 & $s_3 (\times10^{-6})$ & $s_2(\times10^{-3})$ & $s_1(\times10^{-3})$ & $s_0$ \\\hline
$S^{(0)}$ & \llap{$-$}98.96 & 4.625 & \llap{$-$}88.42 & 4.007 \\
$S^{(1)}$ & \llap{$-$}101.6 & 4.750 & \llap{$-$}92.63 & 4.104 \\
$S^{(2)}$ & \llap{$-$}20.83 & 0.9063 & \llap{$-$}8.792 & 3.681 \\
$P_3$ & \llap{$-$}2.604 & 0.1250 & \llap{$-$}4.208 & 0.09700 \\
$P_1$ & 78.13 & \llap{$-$}3.719 & 79.63 & \llap{$-$}0.3260 \\
\hline \hline
\end{tabular}\caption{All three skewness parameters are estimated by the function $S(\lambda)=\sum_{n=0}^3s_n\lambda^n$; this table shows the coefficients $s_n$ for each skewness parameter, as well as the Hermite coefficients of equation (\eqref{eq:genuswnl}).}\label{table:coeffs}
\end{table}
while the value of $\sigma$ can be left as a free parameter.

\subsection{Physical meta-statistics}\label{sec:metastatistics}

Both of the models in Section~\ref{sec:models} can be used as analytic fits to genus measurements from the 2dFGRS data. As the range of scales considered in this work pass well into the non-linear regime, it is to be expected that the genus measurement will deviate substantially from equation (\eqref{eq:genusgrf}).

A measure of such departures was provided by \citet{1992ApJ...392L..51P}, who define the \emph{genus shift} parameter
\begin{equation}
\Delta\nu(\lambda) = \frac{\int_{-1}^{1}\nu g(\nu;\lambda)d\nu}{\int_{-1}^1g_\textrm{model}(\nu;\lambda)d\nu};\label{eq:deltanu}
\end{equation}
when the model accurately reproduces the genus measurement, $\Delta\nu=0$, but when a meatball-- or swiss-cheese--shift are recorded, $\Delta\nu$ becomes greater or less than unity, respectively. The scale-dependence of the parameter is significant because the non-linear effects induced by gravitational evolution cause a modest meatball-shift in the genus curve~\citep{1991ApJ...378..457P}.

To determine which regions of the genus curve account for such deviations---and thus provide motivation for interpretation of the results in terms of voids and clusters---further meta-statistics have been defined by~\citet{2001ApJ...553...33P} and~\citet{2005ApJ...633....1P},
\begin{equation}
A_v(\lambda) = \frac{\int_{-2.2}^{-1.2}g(\nu;\lambda)d\nu}{\int_{-2.2}^{-1.2}g_\textrm{model}(\nu;\lambda)d\nu}\textrm{, and}\label{eq:Av}
\end{equation}
\begin{equation}
A_c(\lambda) = \frac{\int_{1.2}^{2.2}g(\nu;\lambda)d\nu}{\int_{1.2}^{2.2}g_\textrm{model}(\nu;\lambda)d\nu}\label{eq:Ac}
\end{equation}
that probe the deviation from the model in the regions about the troughs of the genus curve below and above $\nu=0$ respectively, in contrast to $\Delta\nu$, which is localised about the central peak.

\section{Results}\label{sec:results}

We measure the genus statistic at 100 points in the range $-4\le\nu\le4$ for the NGP and SGP slices smoothed on scale ranging from 5 to 8 $h^{-1}$Mpc. The errors on the measurements are estimated using the technique of bootstrap resampling~\citep{Efron82,Barrow84}, in which we create distinct samples with the same variance properties as the data by redrawing---with replacement---individual galaxies from the input 2dFGRS catalogue up to the full number of objects; the use of jackknife resampling has not yet been properly explored in the context of the genus measurement. After creating 100 such resampled catalogues, each one is subjected to the method described in Section~\ref{sec:survey}. 

\begin{figure}
 \includegraphics[width=\linewidth]{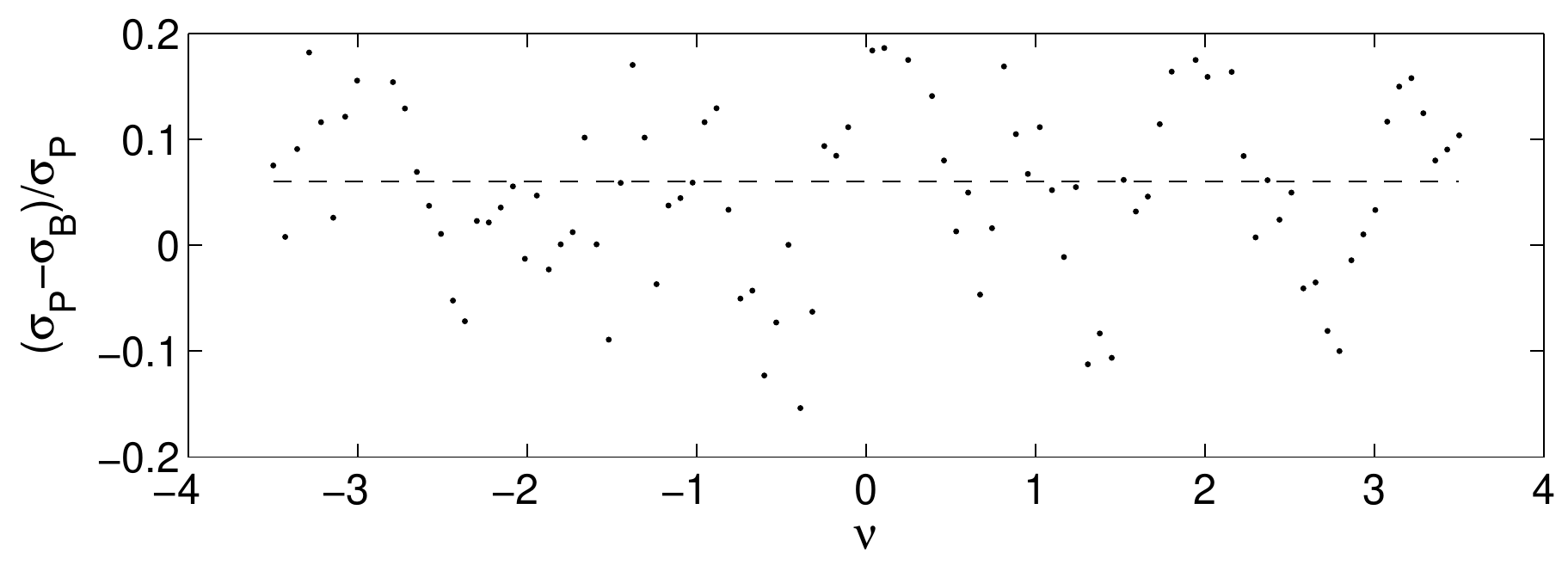}
 \caption{Fractional difference between the bootstrap and Poisson estimates of errors for the genus of a simulated Gaussian random field subject to the same geometric constraints as the data in this work. }\label{fig:bootstrap}
\end{figure}

To test the robustness of the bootstrap resampling, we simulate a Gaussian random density field, with the same geometry as the survey data, from which 100 Poisson samplings are drawn. These are smoothed as \emph{per} the data and the genus of the resulting distributions measured. In Figure \ref{fig:bootstrap}, the variance of these is compared to the variance of the genus measurements from the same number of bootstraps of a single Poisson sample. It is found that, although the bootstrap resamples display on average a meatball-shift, the variance between them is unbiased---on average, the error bars estimated from the bootstrap are 6\% smaller than those from the Poisson samples, but do not exhibit any large-scale $\nu$-dependent trends. Though there is a considerable ($\sim$10\%) tendency for the bootstrap to both over- and under-estimate the error bars in the regions to which $\Delta_\nu$, $A_v$ and $A_c$ are sensitive, this scatter is not systematic; we conclude that the use of bootstrap resampling along with this smoothing method will not create anomolous values for the meta-statistics by virtue of distorting the error bars on the genus measurement.

From the resulting set of mock genus measurements, the covariance matrix of the genus curve is computed for both slices and at each smoothing scale. Recent work~\citep{2007A&A...464..399H} has demonstrated that the bootstrap is a biased estimator of the inverse covariance matrix: the location of the peak in the likelihood function is unchanged---so the best-fit model is correct irrespective---but the slope of the likelihood function in the region about the peak will be overestimated, making the fit appear better than it is; a scaling of the inverse correlation matrix , in accordance with Equation (17) of~\citet{2007A&A...464..399H}, produces an unbiased estimator.

This provides a way to perform a $\chi^2$ analysis of fits to the measurement. For the two models discussed in Section~\ref{sec:models} (Equations~(\ref{eq:genusgrf}) and~(\ref{eq:genuswnl})), we fit the genus measurement by minimising the resulting dispersion 
\begin{equation}
\chi^2 =\frac{1}{n-p} \left[g(\nu) - g_\mathrm{model}(\nu)\right]^\mathrm{T}\mathrm{\mathbf{C}}^{-1}\left[g(\nu) - g_\mathrm{model}(\nu)\right],
\end{equation}
where $\mathbf{C}^{-1}$ is the re-scaled covariance matrix, $n=100$ is the number of data points and $p$ is the number of model parameters and is either one or two.

In this section, we compare the 2dFGRS observations to the models and examine trends with scale, as well as differences arising from the choice of model and meta-statistics; these results are related back to the distributions of voids and clusters.

\subsection{2dFGRS topology below the linear regime}
\begin{figure*}
\centering
\includegraphics[width=\linewidth]{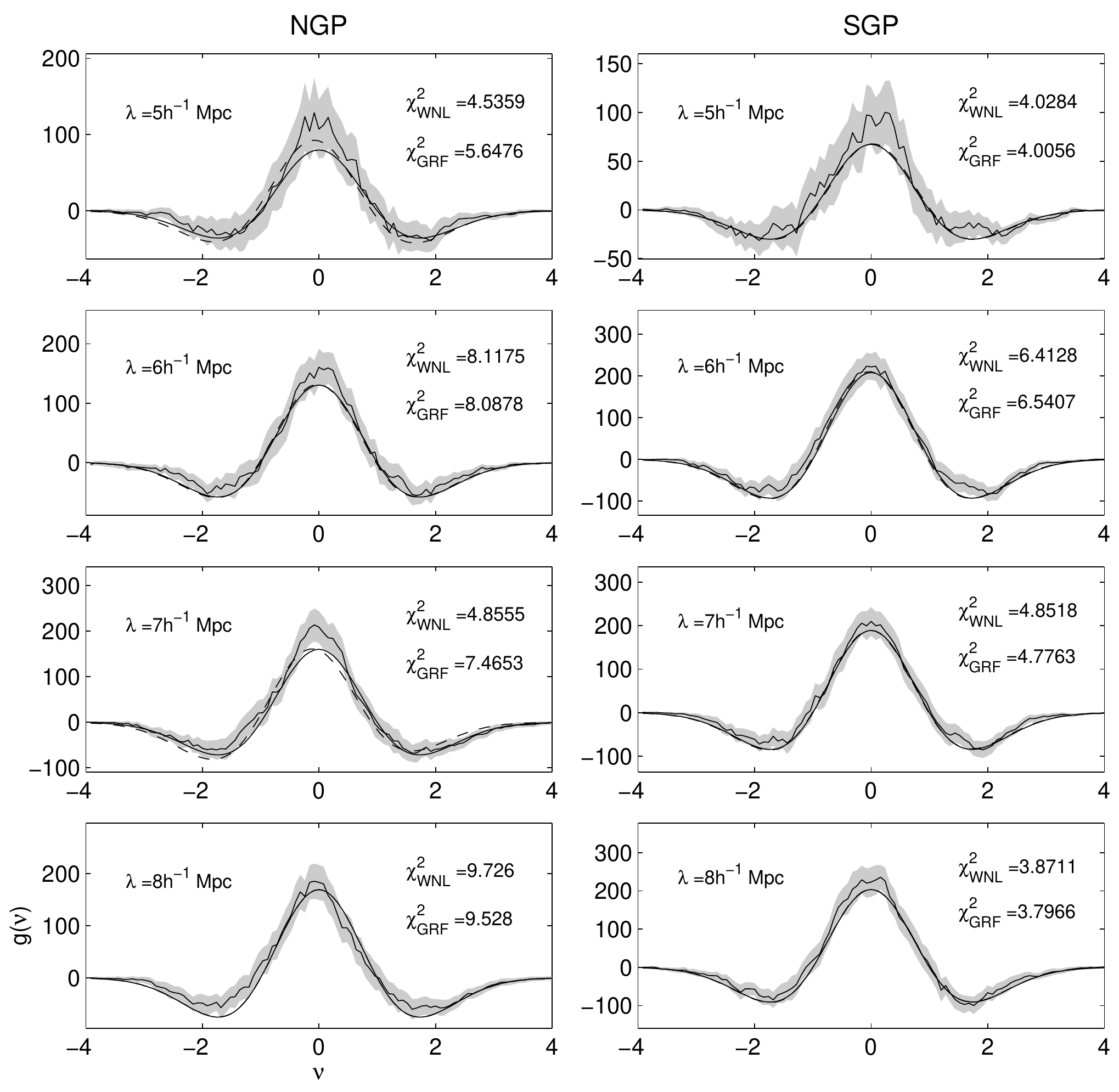}
\caption{Genus curves for samples drawn from the NGP (left) and SGP slices of the 2dFGRS over scales from 5 to 8 $h^{-1}$Mpc, with 3$\sigma$ uncertainty region estimated from the variance between bootstrap resamples of the data sets. The two fits to the measurements are made with the models of Equations~(\ref{eq:genusgrf}) (smooth solid line), leaving the amplitude of the curve as a free parameter, and (\eqref{eq:genuswnl}) (dashed line), with both the amplitude and density rms $\sigma$ varying; in several cases the two curves are nearly identical. The goodness-of-fit values quoted are the reduced $\chi^2$, accounting for the additional parameter in our fit of the weakly non-linear model.}\label{fig:2df_genus}
\end{figure*}

The genus measurements for the NGP and SGP slices are shown in Figure~\ref{fig:2df_genus}, along with the best-fitting Gaussian random field and weakly--perturbed model genus curves.
When calculating statistics such as $\Delta\nu$ the model fitting is done by minimising the scatter in the region $-3\le\nu\le3$, rather than the traditional $-1\le\nu\le1$, in order to use information from all parts of the genus curve. The values of the meta-statistics described in Section~\ref{sec:metastatistics} for the NGP and SGP slices, taken relative to the best-fit genus curve of a Gaussian random field (eq.~\ref{eq:genusgrf}), are given in Table~\ref{table:meta-statistics}.
\begin{table}
\centering
\begin{tabular}{|c|c|c|c|}
\hline \hline
NGP &&& \\\hline
 $\lambda$ & $\Delta\nu$ & $A_c$ & $A_v$ \\\hline
 5 & $0.0063\pm0.0277$ & $0.9957\pm0.1113$ & $0.8534\pm0.0840$ \\
 6 & $0.0158\pm0.0167$ & $0.8056\pm0.0464$ & $0.7400\pm0.0388$ \\
 7 & \llap{$-$}$0.0399\pm0.0166$ & $0.8997\pm0.0508$ & $0.7378\pm0.0433$ \\
 8 & \llap{$-$}$0.0598\pm0.0150$ & $0.7693\pm0.0438$ & $0.6348\pm0.0336$ \\\hline\hline
 SGP &&& \\\hline
 5 & $0.0070\pm0.0230$ & $0.6741\pm0.0576$ & $0.8143\pm0.0670$ \\
 6 & $\llap{$-$}0.0249\pm0.0129$ & $0.7811\pm0.0337$ & $0.7569\pm0.0384$ \\
 7 & $\llap{$-$}0.0060\pm0.0143$ & $0.8850\pm0.0360$ & $0.8054\pm0.0361$ \\
 8 & $\llap{$-$}0.0295\pm0.0148$ & $0.9926\pm0.0412$ & $0.8040\pm0.0367$ \\
\hline \hline
\end{tabular}\caption{Asymmetry meta-statistics for the 2dFGRS samples measured relative to the best-fit genus curve for a Gaussian random field as \emph{per} Equations~(\eqref{eq:deltanu}), (\eqref{eq:Av}) and (\eqref{eq:Ac}). 1$\sigma$ confidence limits are the standard deviation of the set of mock asymmetry statistics generated using the resampled genus measurements, relative to the same model curve.}\label{table:meta-statistics}
\end{table}

These measurements show that for both the NGP and SGP samples, $\Delta\nu$ deviates below zero, indicating that the samples are cluster-dominated at some scales (a meatball-shift topology); we can test the significance of this deviation using the statistic
\begin{equation}
\chi^2_{\Delta\nu} = \frac{1}{4}\sum_{\lambda=5/h}^{8/h} \left(\frac{\Delta\nu(\lambda) - \Delta\nu_\textrm{GRF}}{\sigma_{\Delta\nu}(\lambda)}\right)^2\label{eq:chi2}
\end{equation}
for each of the samples, which measures the deviation from the Gaussian value $\Delta\nu_\textrm{GRF}=0$, and which is tabulated within Table~\ref{tab:tests}. This shows that, assuming the uncertainties are normally distributed, the topology of the SGP region is unlikely to be consistent with a Gaussian random field and that the NGP is definitely not to a high level of significance. This is in agreement with the results of simulations presented by~ \citet{2005ApJ...633....1P} for the case of biased peaks of the $\Lambda$CDM density field, and inconsistent with other cases including unbiased dark matter and haloes.

\begin{table}
\centering
\begin{tabular}{|c|c|c|c|c|c|c|}
\hline \hline
& \multicolumn{3}{|c|}{Against the hypothesis of} & \multicolumn{3}{|c|}{Against the hypothesis}\\ 
& \multicolumn{3}{|c|}{a Gaussian random field} &  \multicolumn{3}{|c|}{of no evolution} \\\hline
  & $\chi^2_{\Delta\nu}$ & $\chi^2_{A_c}$ & $\chi^2_{A_v}$ & $\chi^2_{\Delta\nu}$ & $\chi^2_{A_c}$ & $\chi^2_{A_v}$ \\\hline
 NGP & $5.66$ & $12.3$ & $50.74$ & $3.35$ & $1.57$ & $2.21$\\
 SGP & $2.00$ & $21.13$ & $26.33$ & $0.69$ & $6.60$ & $0.29$ \\
 \hline\hline
\end{tabular}\caption{One-sample (reduced) $\chi^2$ tests on the statistics derived from the genus curve measurements, testing in the first place the case that the density field at these scales is a Gaussian random field (the parameters have their Gaussian values $\{\Delta\nu,A_c,A_v\}=\{0,1,1\}$), and in the second place that the topology of the density field is not evolving with scale (the parameters have their constant best-fit values). }\label{tab:tests}
\end{table}

The measurements also indicate that both $A_c$ and $A_v$ are below unity, showing a lower multiplicity of clusters and voids relative to the case of the Gaussian random field. Table~\ref{tab:tests} again quantifies the departure from the Gaussian case, using the analogous statistics to equation~(\ref{eq:chi2}), where the deviation is now from unity rather than zero. \citet{2005ApJ...633....1P} have demonstrated how the value of $A_v$ might descend below unity through (linearly) biased galaxy formation. It is unclear what general mechanism can drive $A_c$ down as well, though a physical argument is that when the average nearest-neighbour separation of the sites of galaxy formation is shorter than the Gaussian case, at a fixed smoothing scale there will less isolated high-density peaks in the field. This is accompanied by greater connectivity of void regions, resulting in a decline in $A_v$; whether this need always be the case is not clear.

\subsection{Scale-dependent evolution}
\begin{figure*}
\centering
\includegraphics[width=\linewidth]{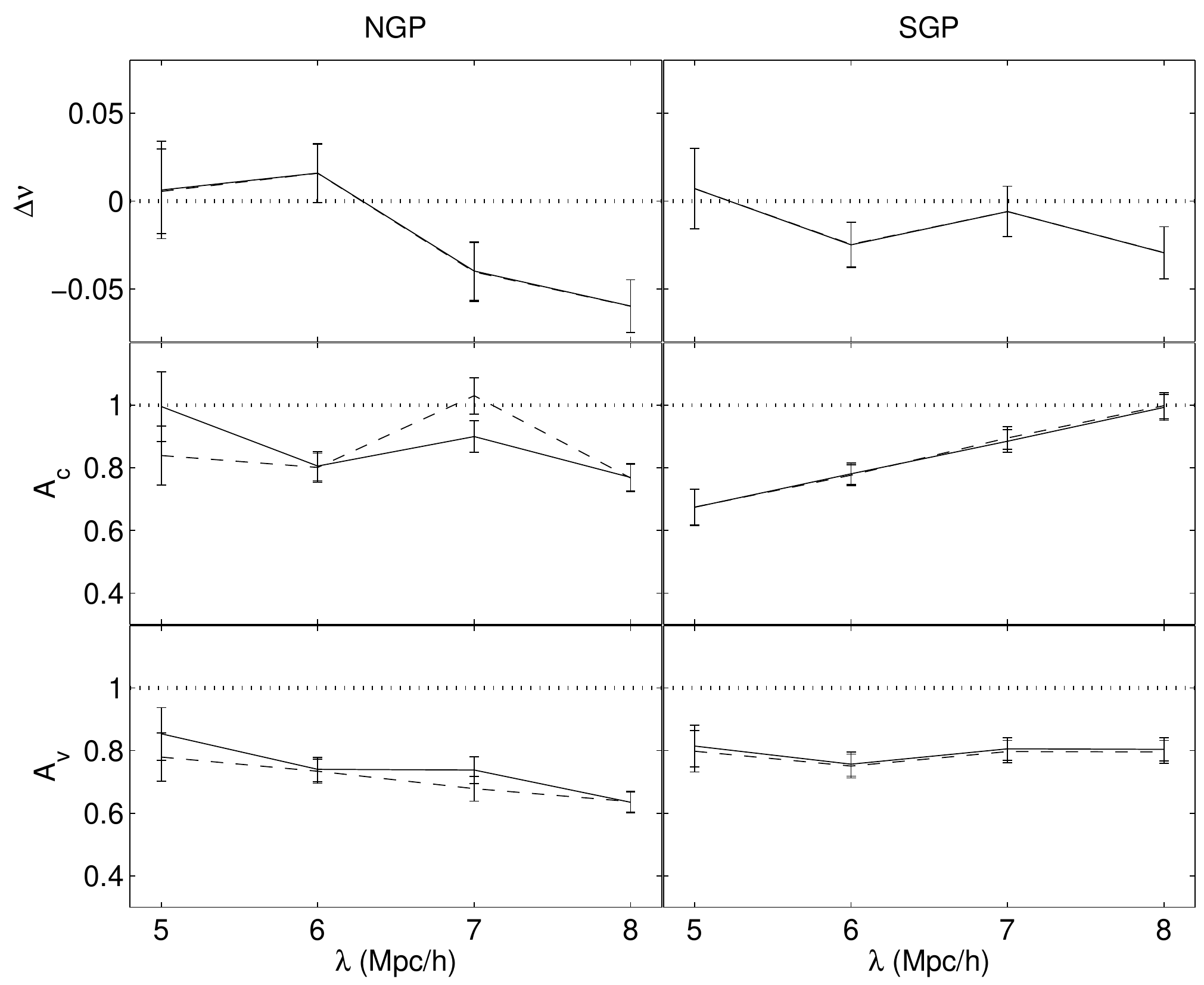}
\caption{Genus curve meta-statistics $\Delta\nu$, $A_c$ and $A_v$ as a function of scale for the 2dFGRS NGP and SGP survey slices, with 1$\sigma$ error bars determined by the variance between the same quantities calculated on each of the bootstrap resamples. Each is calculated by taking the ratio between the data and a model the area under different regions of the genus curve (\emph{n.b.} however a factor of $\nu$ in the integrand of the numerator in Equation~\eqref{eq:deltanu}); results are shown for the calculation relative to the genus curve for a Gaussian random field (solid line) and for its weakly non-linear perturbation (dashed), with the null hypothesis Gaussian random field parameter values (dotted) shown for reference.}\label{fig:asymmetry-broad}
\end{figure*}

There are trends in each of $\Delta\nu$, $A_c$ and $A_v$ as a function of scale shown in Figure~\ref{fig:asymmetry-broad}. The trend in the value of $\Delta\nu$ over the scales examined from negative to zero (within the uncertainty), is in partial agreement with the measurements on simulations by~\citet{2005ApJ...633....1P} who detect a trend from positive to negative as $\lambda$ increases, with a crossing point at $6h^{-1}$ Mpc.
We observe no evidence of such a crossing in the SGP slice, though~\citet{Matsubara96} show that the effect of velocity-space distortions on the small-scale genus curve suppresses the value of $\Delta\nu$ in the weakly non-linear regime. The value of $\Delta\nu$ in the NGP slice more accurately matches the predictions of both theory and simulation, moving from below zero to zero in the range between 7 and 6$h^{-1}$ Mpc.

Table~\ref{tab:tests} also gives the reduced $\chi^2$ for the best-fitting non-evolving values of the parameters. Though the two samples do not agree on trends in $A_c$ and $A_v$, it is clear that the topology is evolving with scale to a high statistical significance. The most remarkable trend is in the SGP slice, which displays a cluster multiplicity consistent with a Gaussian random field on the largest scales and decreasing smoothly to well-below-Gaussian levels at smaller scales. The source of this trend is clearly discernible in Figure~\ref{fig:2df_genus} as a relative elevation in the genus curves over the region in which the cluster multiplicity integral is calculated. Moreover, this elevation broadens and shifts to larger values of $\nu$ on smaller scales, indicating an increasing degree of connection of high-density clusters, which i) occupy an increasingly extreme fraction of the sample volume and ii) extend across a widening range of densities. To complement this, the void multiplicity is essentially stable at sub-Gaussian field levels.

The NGP slice also displays evolution, though the trends are less clear: while the void multiplicity appears to increase toward smaller scales, this could be a result of suppression in the number of isolated voids at \emph{larger} scales due to large, connected over-dense structures within the NGP field. It is possible to test for the two samples describing the same evolution in each parameter using the statistic
\begin{equation}
\chi^2_\textrm{cross} = \frac{1}{4}\sum_{\lambda=5/h}^{8/h}\left(\frac{\Delta\nu_\textrm{NGP}(\lambda) - \Delta\nu_\textrm{SGP}(\lambda)}{\sqrt{\sigma_\textrm{NGP}(\lambda)^2 + \sigma_\textrm{SGP}(\lambda)^2}}\right)^2,
\end{equation}
where each $\sigma$ is the error bar on the $\Delta\nu$ parameter and where analogous statistics can be defined for $A_c$ and $A_v$. The NGP and SGP samples do not show consistent trends and offsets in the parameter values, with the cross-statistic yielding 2.05 ($\Delta\nu$), 5.16 ($A_c$) and 3.30 ($A_v$). On the scales probed in this work, it appears that the topology of structure differs modestly between the two regions at median density and more strongly for over- and under-dense structures.

\subsection{Effect of model choice}
A comparison of goodness-of-fit between the Gaussian and weakly non-linear models for the genus curve does not provide conclusive evidence that the weakly non-linear curve is superior. In most cases, the difference between the models is negligible, though when it is not, the weakly non-linear curve is much to be preferred. However, the strongest potential confirmation of the weakly non-linear theory---that the quality of the fit of this model improves relative to the Gaussian one on progressively smaller scales---cannot be demonstrated based on these results. This reinforces the claim that the effect on the genus curve of accounting for weak departures from Gaussianity is modest, though one must expect this will not hold as such departures grow in amplitude. 

The results could be taken to indicate that this effect will be best seen on scales smaller than those that have been probed and requires somewhat larger volumes than are available on these scales with the present data. An interesting alternative is to question the impact of differing forms of galaxy bias on the genus curve. The 2dFGRS has been used (with statistics other than the genus curve) to study the linear and quadratic bias of galaxies~\citep{2002MNRAS.335..432V} as well as the stochasticity of the relative bias between early- and late-type galaxies~\citep{2005MNRAS.356..247W}; topological statistics may be sensitive to a third characterisation of bias as \emph{non-local}, as the smoothed distribution incorporates information from regions around the sites of individual galaxies. These will certainly lead to asymmetries in the genus curve of the kind observed here, though it is not clear whether these will involve different modifications to those incorporated in the formalism of equation~(\ref{eq:genuswnl}). 

It is also possible to investigate the effect of using the weakly non-linear genus curve model when measuring the parameters $\Delta\nu$, $A_c$ and $A_v$. The dashed lines in Figure~\ref{fig:asymmetry-broad} show the value of these parameters when measured relative to the weakly non-linear genus curve with skewness parameters selected as a function of scale. No distinction is to be made between the models in the SGP slice, while the values of the parameters in the NGP slice differ at those scales where the weakly non-linear model provides a substantially better fit. 

\subsection{Comparison with larger scales}
\begin{figure*}
\centering
\includegraphics[width=\linewidth]{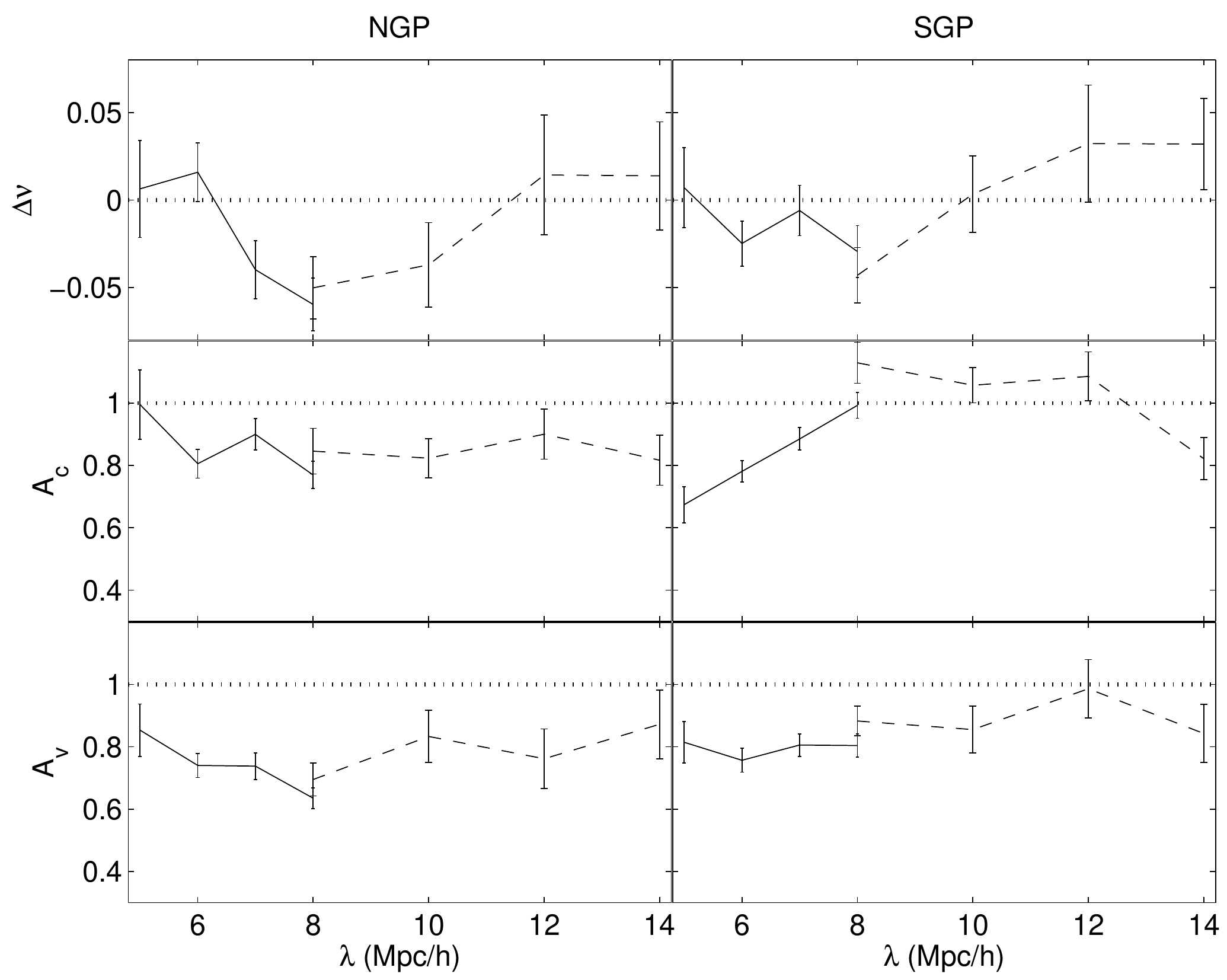}
\caption{Comparison of meta-statistics $\Delta\nu$, $A_c$ and $A_v$ for the  2dFGRS NGP and SGP survey slices between the scales studied in this paper (solid lines; identical to Figure~\ref{fig:asymmetry-broad}) and those of~\citet{2007MNRAS.375..128J} (dashed lines).}\label{fig:asymmetry-comp}
\end{figure*}

\citet{2007MNRAS.375..128J} have studied the topology of large-scale structure in the 2dFGRS on scales ranging from 8 to 14 $h^{-1}$Mpc, using slightly different methodology to that employed here with regard to the construction of statistically complete samples, as well as in the parameterisation of asymmetries in the genus curve. Figure~\ref{fig:asymmetry-comp} extends the results in the present work to larger scales by calculating $\Delta\nu$, $A_c$ and $A_v$ for the earlier results.

The two sets of results are consistent at the overlapping scale of 8 $h^{-1}$Mpc, particularly considering the methodological differences. The cluster parameter in the SGP is a possible exception to this, and indeed it seems clear that little inference about the scale-dependence of this statistic across the full range is possible given the data presented here. The trends that do emerge are: i) the shift parameter is consistent with zero at larger scales, drops below zero at around 10 $h^{-1}$Mpc and returns to at the smallest scales studied in this work; ii) the void parameter changes remarkably little when moving in to the non-linear regime and is consistent with $A_v\approx0.8$ across all scales and between both regions of the survey. This value for the void parameter is also seen in the topology of the Sloan sub-regions reported by~\citet{2006astro.ph.10762G}, as well as their measurements on the Millennium Run~\citep{2005Natur.435..629S} and on simulations of dark matter haloes populated with galaxies~\citep{2005ApJ...633....1P,2006ApJ...639..600K}, to the point where it seems to deserve recognition as a generic feature of genus measurements. 

Further work in this area should aim to characterise the scale at which the shift parameter departs from zero and, if the turn-around at lower scales is reproduced in independent samples, the scale at which the meatball-shift is maximised. Though the void parameter is now routinely observed to depart from unity, no complete explanation for this observation has emerged. The understanding of both of these trends would profit from examination of the differences between particular galaxy populations. The continued progress of the SDSS and the early results reported by~\citep{2005ApJ...633...11P} auger well for this effort. Forthcoming surveys based on photometric redshifts are likely to make a substantial contribution through the two-dimensional genus statistic in redshift slices. 

It is interesting to speculate on the correspondence between the angular- and three-dimensional correlation functions and the two- and three-dimensional genus statistics, as well as between the ratioing process of~\citet{1993ApJS...86....1M} and the common practice of random point catalogues to account for geometric and astronomical selection effects in the calculation of correlation functions. Progress on the interpretation of the genus measurements may also require the use of alternative meta-statistics. A possible way forward is to use the orthogonality relations of the Hermite polynomials and decompose the genus curve into a series of Hermite functions $\psi_n(\nu) \propto \exp(-\nu^2/2)H_n(\nu)$. The co\"efficients of the decomposition are linked to the skewness parameters of the field---in the linear regime only the $n=2$ co\"efficient is non-zero, while in the non-linear regime the $n=1$ and $n=3$ functions contribute with some small amplitude to the genus curve, inducing a slight asymmetry. A full decomposition of the curve on scales at and below those studied in this work would guide theoretical interpretation and prediction of the topology of non-linear structure.

\section{Conclusions}\label{sec:conclusions}

We have devised a method for measuring the genus curve on flux-limited samples from galaxy redshift surveys in the non-linear regime, setting out the requirements for avoiding major systematic effects in the form of a joint constraint diagram. This method has been applied to the 2dF Galaxy Redshift Survey, providing a smoothed galaxy distribution on which the topology of large-scale structure in the weakly non-linear regime, from 5 to 8 $h^{-1}$Mpc, has been studied. These measurements are the shortest-scale study of structure topology carried out to date with observational data and have been aimed explicitly at detecting the effects of non-linear structure formation on the genus statistic.

By measuring the genus shift, void multiplicity and cluster multiplicity statistics of the genus curve, we have aimed to quantify departures from the linear regime case of a Gaussian random field. We have observed the effects of structure formation on the genus curve in the form of non--linear gravitational evolution, as demonstrated by a trend in genus shift, and demonstrated that both the cluster and void multiplicity statistics tend to fall below unity across a broad range of scales. The independence with scale of the latter quantity in particular suggests effects external to the non-linear evolution of the dark matter field, in agreement with earlier suggestions about a relationship between $A_v$ and galaxy bias. 

We have studied the effect of model choice in the measurement of these statistics, using analytic predictions for both the linear regime and weakly non-linear regime as comparators to the data. We find that neither model adequately captures the shape of the genus curve, though there is evidence at some scales that the weakly non-linear model provides a substantially better fit. Together, these measurements demonstrate the efficacy of studying topology in the non-linear regime and, given the emphasis future surveys are placing on higher-order statistics, suggest that topology will play an important r\^ole in answering questions about the formation and evolution of galaxy clusters and dark matter haloes on these scales.

\section*{Acknowledgments}
A significant part of the research presented in this paper was conducted as a thesis for the completion of a Master of Science degree at the University of Sydney. This research was supported by an University Postgraduate Award and a Denison Postrgraduate Award as well as by the generous assistance of St Andrew's College within the University of Sydney. JBJ would like to acknowledge the support shown by members of the Institute of Astronomy at the University of Sydney, the Anglo-Australian Observatory and at the Institute for Astronomy within the University of Edinburgh. I.~P.~R.~Norberg and  A.~F.~Heavens contributed helpful criticisms throughout the preparation of this paper.
\bibliographystyle{mn2e}
\bibliography{ms}

\appendix

\bsp

\label{lastpage}

\end{document}